*Article*

# Genetic Information as a "Chord" of Chemical Oscillations: Emergence of Catalyst-RNA Systems Driven by Superposed Rhythms

**Takeshi Ishida [1]***

[1] Department of Ocean Mechanical Engineering, National Fisheries University, Shimonoseki 759-6595, Japan; ishida@fish-u.ac.jp
* Correspondence: ishida@fish-u.ac.jp

**Abstract**

A central challenge in the origin of life is understanding how catalytic peptide-like polymers and information-bearing nucleic acid-like polymers emerged as an interdependent system. This study constructs a "primordial cognitive model" incorporating two internal Lotka-Volterra chemical oscillators to investigate, through simulation, whether a catalytic loop, which comprises functional peptides, primordial tRNAs, and nucleic acids that record and amplify them, can form through the interaction of polymers represented by binary (0/1) sequences. In this model, a mechanism was introduced where the synthesis of internal oscillations provides a temporal bias for 0/1 selection during polymer elongation, while generated functional sequences are protected, recorded, and re-amplified. Simulation results demonstrated that the proposed cognitive model significantly outperformed a contrast model based on random 0/1 selection in terms of the establishment rate of catalytic loops, the accumulation of functional molecules, polymer elongation, and the reduction of Shannon entropy in sequence distribution. Furthermore, this superiority was generally maintained across sensitivity analyses, including batch calculations with different random seeds, changes in target sequences, and variations in thresholds, substrate adjustment units, catalyst and tRNA tolerances, protocell numbers, and fitness weights. These findings suggest that temporal bias based on internal chemical oscillations effectively narrows the search within vast sequence spaces, facilitating the emergence and retention of functional sequences. While this study is a computational model based on abstract binary sequences and simplified translation/replication rules rather than a direct reconstruction of life’s origin, it provides a working hypothesis for the interdependent emergence of catalytic function and information retention by demonstrating that internal oscillations can bias sequence exploration within a framework linking autocatalytic networks, recording, and group selection. Future research must further verify the generality and empirical validity of this framework by expanding monomer types, evolving into multi-oscillator systems, and establishing correspondences with compartmentalized experimental systems.

# 1. Introduction

*1.1. Emergence of the Primordial "Catalyst-RNA System"*

Various theories concerning the origin of life have been proposed, including the RNA world hypothesis, where RNA initially performed both genetic and catalytic functions, and the peptide-RNA co-evolution theory, which posits early interactions between proteins and RNA. The RNA world hypothesis, fundamentally outlined by Gilbert [1], is supported by the discovery of self-splicing RNA and catalytic RNA in ribonuclease P [2-4]. These findings demonstrate that RNA can function as an enzyme, supporting its central role in prebiotic stages. Conversely, recent reviews [5,6] re-evaluate peptide-RNA co-evolution, suggesting that since modern gene expression relies on the deep interdependence of RNA and proteins, they likely co-evolved by expanding their respective functional roles in parallel. From this perspective, treating the interdependent emergence of catalytic and information molecules within a single theoretical model is of significant importance.

Regarding the environmental setting of life's origin, numerous hypotheses such as hydrothermal vents and hot springs have been suggested. The GADV hypothesis proposed by Kenji Ikehara [7, 8] posits a protein-led origin, emphasizing the central role of four amino acids: glycine (G), alanine (A), aspartic acid (D), and valine (V). This theory suggests that globular "GADV proteins" composed of amino acids corresponding to GNC codons (G: guanine, N: any, C: cytosine) initially exhibited weak catalytic functions, establishing the foundations for self-replication and metabolism. In this scenario, primordial tRNA and modern coding systems gradually evolved from these beginnings into complex biological systems.

Two prominent views exist regarding the origin of tRNA and the genetic code: the operational RNA code hypothesis [9], based on recognition by the tRNA acceptor stem, and the stereochemical hypothesis [10], based on the direct affinity between RNA and amino acids [11]. The "primordial tRNA" discussed in this study represents an abstraction of the minimal unit forming the correspondence between amino acids and short code sequences, rather than a reconstruction of the modern tRNA molecule. Furthermore, the possibility that RNA mediated aminoacylation of tRNA-like molecules is supported by research on aminoacylation ribozymes and flexizyme systems. Recent studies have identified ribozymes derived from natural tRNA-sensing systems that possess self-aminoacylation activity, suggesting that bridging mechanisms could have been established by RNA prior to the evolution of translation [12-13].

However, these hypotheses fail to sufficiently describe the dynamic process by which catalytic peptides and information-bearing nucleic acids emerge as an interdependent system within a single computational framework. The "chicken-and-egg" problem, where nucleic acids require peptides for replication and peptides require nucleic acids for information, remains a central challenge. Functional sequences are extremely rare in the vast sequence space; thus, their accumulation through random polymerization and degradation is highly improbable. Furthermore, as Eigen noted, an "error catastrophe" occurs when the mutation rate exceeds a certain threshold, preventing the maintenance of meaningful genetic information [14]. Overcoming this combinatorial explosion requires a selective mechanism that guides monomers toward specific sequences.

To address this, two chemical oscillators are introduced as a candidate selective mechanism. By defining functional molecules that bridge two types of monomers, acting as primordial tRNAs, this study investigates whether an interdependent system of catalytic sequences and recorded nucleic acids emerges more effectively than in random controls. While the use of chemical oscillators may seem unconventional, it follows the logic of Turing's pattern formation: just as spatial interactions between oscillations generate self-organized forms, temporal interactions can facilitate the emergence of information-processing structures. This approach aligns with dissipative structure theory [15], where non-equilibrium oscillations generate order by consuming energy, and relates to hypercycle theory [16], which stabilizes the coexistence of information and function through cyclic coupling.

Previous work has proposed models for the emergence of cell morphology [17] and homeostasis in geneless protocells [18] driven by chemical oscillators. The latter represents a minimal implementation of "active inference" [19], where a system maintains its existence by predicting and adapting to its environment. In that model, a Lotka-Volterra oscillator served as a "search engine" to fluctuate the phenotype, while a feedback loop optimized the state based on its correlation with a global success metric. The present study extends this cognitive process from mere homeostasis to the emergence of catalytic networks with nucleic acid recording, investigating how chemical oscillators can "infer" and select meaningful sequences.

### *1.2. Overview of Related Previous Research*

To overcome the combinatorial explosion problem, various candidate selective mechanisms have been proposed to guide monomers toward specific sequences. These include inorganic catalysts such as metals and minerals, physicochemical affinities between molecules, and compartmentalization via liquid-liquid phase separation (LLPS).

Regarding the emergence of functional molecules from inorganic catalysts, scenarios involving the co-evolution of cofactors have been suggested [20], where inorganic substances initially acted as catalysts, with synthesized organic molecules subsequently evolving into more advanced catalysts. Specifically, within the lineage of surface metabolism and the "iron-sulfur world", treating the surfaces of iron-sulfur minerals or Ni/Fe sulfides as reaction sites, the possibility that mineral surfaces supported both early metabolism and organic molecule synthesis has long been discussed. Experimental evidence demonstrating peptide formation from amino acids on Ni/Fe sulfide surfaces under CO conditions supports the role of inorganic surfaces as scaffolds for functional polymer generation [21-22].

In recent years, protocells based on LLPS, such as coacervates, are being re-conceptualized not merely as "concentration sites" but as platforms for "chemical memory" capable of recording internal states and stimulus-responsive information processing [23]. Furthermore, experimental systems demonstrating life-like behaviors such as growth and division have been reported, concretizing the potential for compartmentalization to facilitate the maintenance and updating of information [24].

The GADV hypothesis proposed by Ikehara provides a detailed framework for processes such as the emergence of the first catalysts (weakly globular proteins) and the establishment of cellular compartments by peptides [7-8].

The theory of autocatalytic networks proposed by Stuart Kauffman offers a compelling alternative to RNA-first hypotheses. This theory posits collective self-replication, where the network as a whole facilitates the production of its components, rather than individual macromolecules replicating themselves [25-26]. The Reflexively Autocatalytic and Food-generated (RAF) theory defines the mathematical conditions under which a set of all components can be generated from food sources (monomers) through a series of catalyzed reactions [27-28].

While these studies have played a significant role in understanding the formation of prebiotic "hardware," the full scope of how the "software" and "memory devices", the information-processing capabilities mediated by proteins and RNA (or DNA) that allow life to respond to the environment and regulate its own state, emerged remains unresolved.

To elucidate aspects of information processing in primordial cells, a "primordial cognitive model" for the emergence of homeostasis was previously proposed [18]. This model can be regarded as a minimal implementation of the Free Energy Principle, specifically the concept of active inference proposed by Karl Friston [19], which posits that a system maintains its existence by predicting and adapting to its environment and internal states. That research introduced a chemically plausible minimal mathematical model where homeostasis emerges self-organized from a simple reaction network without external genetic information or predefined targets. In the model, an internal Lotka-Volterra oscillator served as a "search engine" to periodically fluctuate the system's phenotype (e.g., cellular pigment). By evaluating only the temporal correlation between these fluctuations and a single global metric of success (reproduction rate), the system adjusted its internal state. This simple loop of "exploration via oscillation and optimization via correlation" was realized through a chemically implementable mechanism involving "antagonistic memory molecules." Simulations demonstrated that the model could autonomously converge toward and maintain an optimal temperature for reproduction despite significant environmental fluctuations. However, the model did not address how functions such as RNA and catalysis emerge from this framework. The present study expands this cognitive process to the emergence of catalytic networks with nucleic acid recording, investigating how chemical oscillators can "infer" and select meaningful sequences.

The idea of chemical oscillators controlling polymerization has gained attention in synthetic chemistry. For instance, studies using the Belousov-Zhabotinsky (BZ) reaction have demonstrated autonomous nanostructure formation and polymerization induction. In BZ-driven self-assembly, oscillating radical production controls the initiation and termination of polymerization, enabling precise control over polymer morphology and cargo encapsulation [29]. Furthermore, reports on supramolecular oscillators describe dynamic supramolecular polymers, more closely resembling biological systems, where redox cycles are coupled with nucleation, elongation, and fragmentation [30-31]. These studies experimentally support the possibility that complex structures and functions can emerge solely from internal rhythms without external precision control [32-33].

Regarding the birth of macromolecules like RNA, environments involving wet-dry cycles, typical of terrestrial hot springs, have been proposed to physically drive prebiotic elongation through dehydration synthesis and rehydration-induced rearrangement/selection [34]. Experimental evidence showing that nucleotide monomers polymerize into long chains through wet-dry cycles reinforces the empirical basis for periodic environments lowering the barriers to polymerization [35]. Given that periodic external driving forces, such as wet-dry cycles, could have been internalized into autonomous oscillations during evolution, information control via oscillators is a natural progression of this concept.

In summary, while previous research has individually addressed compartmentalization, autocatalysis, periodic driving, or information retention, few studies have examined the integrated closed loop in which (i) internal oscillations bias sequence generation, (ii) functional sequences are retained as nucleic acid records, and (iii) those records contribute to the subsequent amplification of functional molecules. Therefore, this study expands the primordial cognitive model [18] to construct an abstract model integrating exploratory bias via internal oscillations with the recording, protection, and amplification of functional molecules.

### *1.3. Overview of the Proposed Model*

Building upon the "primordial cognitive model" [18], this paper proposes a "gene-catalyst emergence model" designed to facilitate the emergence of functional macromolecules (informational molecules) such as RNA and their elongation. Specifically, based on a simulation model consisting of virtual one-dimensional polymers and catalytic reaction groups, this model demonstrates the possibility of preferentially increasing specific sequences with catalytic functions by utilizing two chemical oscillators as a "cognitive engine." These oscillations are modeled using the Lotka-Volterra equations, as in the previous study [18]. Figure 1 provides an overview of the model, which is described in detail in Section 2. The summary of the Lotka-Volterra chemical oscillation model is provided in Appendix A.1.

While the basic model investigates the preferential expression of specific polymer sequences using two oscillators, the superposition of additional oscillators is expected to enable the generation of increasingly complex sequences involving diverse monomers. This approach draws an analogy with "complex pattern generation through wave interference," hypothesizing that complex informational sequences can be expressed through the harmony of multiple oscillations. In other words, this model embodies the philosophy that "genetic information began as a 'chord' of chemical oscillations, which eventually crystallized into information." Just as a superposition of simple sine waves can represent any waveform, the interference of multiple chemical oscillators with different periods generates complex and specific timings for chemical reactions (temporal patterns) within a protocell, which are then transcribed into polymer sequence information (spatial patterns). Although this paper focuses on the simplest case involving two oscillators and the adjustment of oscillation centers, this framework is extensible to multi-oscillator systems.

The novelty of this research lies in: (i) providing an intrinsic bias to the 0/1 selection during polymer elongation through the synthesis of multiple chemical oscillations; (ii) integrating a mechanism to protect, record, and re-amplify generated functional sequences within a single model; and (iii) quantitatively evaluating these effects in comparison to a random control model.

Simulation results indicate that the proposed model outperforms the random control model, which proceeds solely through random reactions, in terms of the frequency of functional sequence generation, the establishment rate of catalytic loops, and the maintenance and increase of polymer length. These findings suggest that the exploratory bias based on internal oscillations effectively narrows the sequence search space, facilitating the accumulation of functional molecules.

Importantly, this model is not intended to replace the RNA world hypothesis, the peptide-RNA co-evolution theory, autocatalytic networks, or the compartmentalization hypothesis. Instead, it serves as a computational working hypothesis to bridge these theories by providing a framework to investigate how internal oscillations can mediate sequence search and information retention.

Regarding related studies, the compositional genome (Graded Autocatalysis Replication Domain (GARD) model) has been proposed as a comparative model for "pre-genetic" inheritance that does not rely on sequence (digital) information, where the composition of amphiphilic molecular assemblies is autocatalytically maintained and replicated [36]. While the GARD model provides compositional inheritance through "compartmentalization + mutual catalysis," the present proposal differs in its use of periodic driving (oscillators) to narrow the search space and accelerate the generation of "sequence-based informational molecules." This contrast helps clarify the advantages of the current model, such as elongation and the overcoming of the error catastrophe.

A defining characteristic of this model is that the accumulation of functional molecules arises solely from the interaction between internal reactions and internal oscillations, without directly providing specific sequences or target values from outside the protocell. Furthermore, this study quantitatively tracks the progression of internal information localization using the Shannon entropy of sequence distribution as an indicator. In the framework of information thermodynamics, mutual information and entropy are organized as quantitative indices linking "knowledge (memory)" and "dissipation (irreversibility)" [37]. Therefore, tracking the temporal evolution of internal information entropy in this study is considered a valid approach to positioning information accumulation within chemical reaction networks as a physical quantity. The calculation procedure for information entropy in this model is detailed in Appendix A-2.

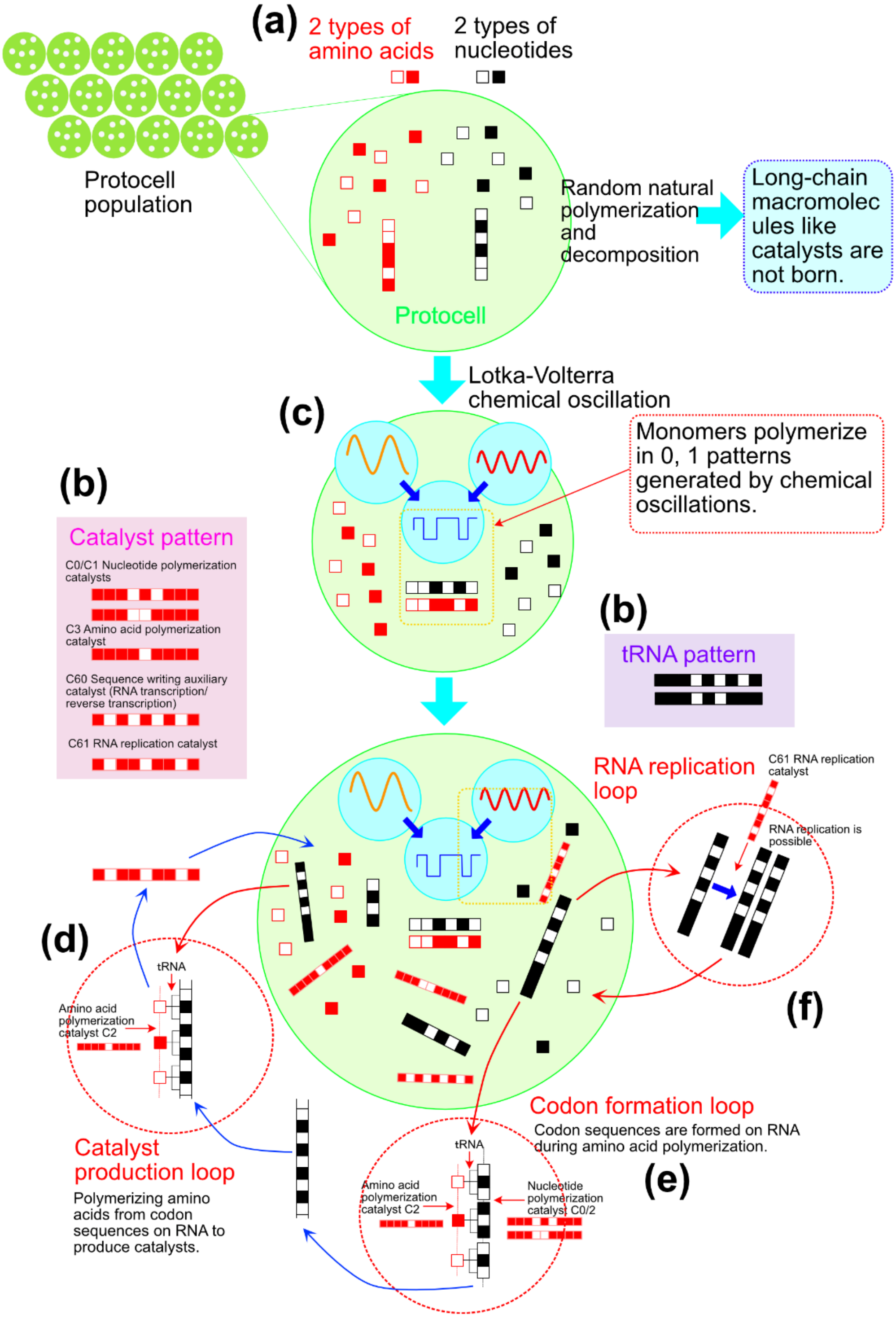

**Figure 1. Schematic overview of the model.** A population of virtual protocells (e.g., 50 cells) evolves based on fitness. Each protocell contains two types of virtual monomers. While random natural polymerization and decomposition rarely produce meaningful sequences, the introduction of two internal chemical oscillators and polymerization that probabilistically synchronizes with these oscillations enables the preferential production of meaningful sequences such as catalysts. When primordial tRNAs, which bridge specific amino acids with three-digit nucleotides, and five types of catalysts are present, catalytic loops for RNA replication and RNA-mediated catalyst production are established.

## 2. Materials and Methods

### *2.1. Basic Configuration and Computational Space*

This study utilizes a population model of protocells consisting of virtual nucleic acid and peptide chains represented by binary symbols (0 and 1) (Fig. 1(a)). Each protocell operates as an independent reaction system, and multiple protocells are simulated in parallel under identical conditions. The number of protocells of reference parameters set is set to 50. Each simulation runs for 2,000 steps, with a numerical integration step of dt = 0.08 for the Lotka-Volterra oscillators.

Equal amounts of nucleic acid monomers (0 and 1) and amino acid monomers (0 and 1) are initially allocated within each protocell. Under reference parameter conditions, the total number of monomers for both nucleic acids and amino acids is 50,000 each. The initial population includes 100 random nucleic acid chains and 100 random peptide chains with lengths ranging from 1 to 10. The maximum polymer length is limited to 100, and the capacity of the natural polymerization pool is capped at 100 chains for both nucleic acids and peptides. If the number of chains in the natural polymerization/decomposition pool falls below 30, corresponding 1-bit monomers are automatically replenished. This configuration coarse-grains the continuous supply of short molecules in a prebiotic environment.

In this model, both nucleic acid and peptide chains are represented as 0/1 sequences. Nucleic acids serve as information carriers for replication, while peptides exhibit catalytic activity upon reaching specific sequences. It should be noted that the "nucleic acids" and "peptides" described here are not literal representations of modern RNA or proteins but are simplified abstract representations of primordial informational and functional macromolecules. Thus, while these terms are used for convenience, they fundamentally represent virtual polymers with binary sequences.

Natural polymerization and decomposition are applied to each polymer chain as default reactions. Natural polymerization involves adding a single bit to the end of each chain at each step, with a reference rate of 0.04. Natural decomposition is modeled as the removal of a single bit from a random position within the chain, rather than terminal cleavage. The initial decomposition rate is 0.14, increasing by 0.000002 per step. This setup coarse-grains the strengthening of degradation-promoting effects, not explicitly modeled, as the reaction system increases in complexity.

### *2.2. Basic Configuration and Computational Space*

Two types of nucleic acid sequences are defined as primordial tRNAs [Fig. 1(b)]. tRNA-1 (sequence: 101110<u>101</u>) facilitates the correspondence between codon 100 and amino acid 0, while tRNA-2 (sequence: 101110<u>111</u>) facilitates the correspondence between codon 110 and amino acid 1. Here, 100 and 110 serve as 3-bit codons that record the 0-bit and 1-bit of a peptide, respectively. By mapping each bit of a peptide sequence to a 3-bit codon, functional peptide sequences can be recorded on the nucleic acid side [Fig. 1(e)].

Five types of 9-bit catalytic peptide sequences are defined [Fig. 1(b)]: 111<u>010</u>111 (c0) promotes nucleic acid 0 polymerization; 111<u>001</u>111 (c1) promotes nucleic acid 1 polymerization; 111<u>101</u>111 (c3) promotes amino acid polymerization; <u>101010101</u> (c60) aids the process of transcribing peptide sequences into nucleic acid codons; and 101101101 (c61) replicates protected nucleic acid records. These correspond to two types of nucleic acid polymerization catalysts, one amino acid polymerization catalyst, one sequence-writing auxiliary catalyst, and one nucleic acid replication catalyst, respectively.

The recognition of functional molecules is performed by evaluating the terminal 9 bits of newly elongated chains, rather than scanning the entire length of all chains in the natural polymerization/decomposition pool. Specifically, immediately after a peptide or nucleic acid chain elongates, if its terminal 9 bits match a predefined functional sequence, this 9-bit segment is excised from the natural pool and transferred to a "protected functional molecule pool." This protected pool contains catalytic peptides, primordial tRNAs, and nucleic acid molecules that record catalytic sequences as 3-bit codon strings. Consequently, the model maintains a two-layer structure consisting of free chains in the natural pool and stabilized functional molecules in the protected pool.

Protected functional molecules are treated as more resistant to decomposition than free chains. While free chains degrade at the current natural decomposition rate, molecules in the protected pool are designed to decrease at only 5% of that rate. This two-layer representation models the relative stabilization of functional molecules through folding or complex formation.

In addition to exact sequence matching, weak functional activity based on Hamming distance is permitted. For peptides, a 9-bit exact match provides a relative activity of 1.0, while sequences with Hamming distances of 1 and 2 are treated as weak catalysts with weights of 0.1 and 0.05, respectively (Hamming distance is detailed in Appendix C). For tRNAs, sequences differing by 1 bit are also allowed as weak functional molecules. This allowance enables neighboring sequences with weak functions to support the initiation of the reaction network, even when exact matches are extremely rare in the early stages.

*2.3. Configuration of Binary (0/1) Selection via Dual Chemical Oscillators*

Each protocell incorporates two internal Lotka-Volterra chemical oscillators [Fig. 1(c)]. The waveform of each oscillator is determined by the substrate amount $G$ (represented as $A$ in Appendix A), the concentrations of two internal chemical species $X$ and $Y$, and three reaction rate constants $k_1$, $k_2$, and $k_3$. Oscillator A generates a long-period, large-amplitude wave, while Oscillator B generates a short-period, small-amplitude wave. Under reference conditions, Oscillator A was set with $G$ = 1.0, initial concentrations ($X$, $Y$) = (9, 10), and rate constants (0.8, 0.1, 0.6). Oscillator B was set with $G$ = 0.5, initial concentrations (9, 10), and rate constants (1.5, 0.1, 0.9). These initial values yield near-sinusoidal waveforms with distinct periods (refer to Appendix E for waveform profiles). Additionally, small random fluctuations were applied to the initial concentrations in each protocell to create a population with slight variations in phase and amplitude despite identical parameters.

For each oscillator $i \in \{A, B\}$, the temporal evolution of the concentrations $X_i$ and $Y_i$ is described by Lotka-Volterra equations. In this model, $Y_i$ is regarded as the output-side species, and the product amount $P_i(t)$ is defined as:

$$P_i(t) = k_i Y_i(t)$$

where $k_i$ is a scaling coefficient. By synthesizing the products $P_i(t)$ from the two oscillators, the decision variable $D(t)$ for 0/1 selection is defined as:

$$D(t) = P_A(t) - P_B(t) - \theta$$

Differential synthesis was employed with a reference threshold $\theta = -6$. The fundamental rule for the cognitive model is to select 1 if $D(t) > 0$ and 0 if $D(t) \leq 0$. In contrast, the random model selects 0 or 1 with equal probability, independent of the oscillators. Consequently, in the cognitive model, $D(t)$ fluctuates over time through the superposition of the two oscillators, generating a sequence for 0/1 selection during polymer elongation.

This binary selection was applied during both natural and catalytic polymerization. In the cognitive model, rather than appending the 0 or 1 obtained from oscillator synthesis directly, a mutation rate of 0.001 was introduced to allow for the stochastic flipping of bits. This mechanism ensures that while the chemical oscillations strongly guide sequence generation, the process remains exploratory and avoids completely deterministic sequence fixation.

### *2.4. Configuration of the Polymer Reaction System: Polymerization, Translation, Recording, and Replication*

Catalytic acceleration was integrated into the polymerization process of free chains. The elongation probability for nucleic acid chains was defined as the sum of the natural polymerization rate (0.04) and an increment proportional to the combined activity of nucleic acid polymerization catalysts (c0 and c1). Similarly, the elongation probability for peptide chains was defined as the natural rate (0.04) plus an increment proportional to the activity of the amino acid polymerization catalyst (c3). In the reference parameters, the acceleration coefficients for both nucleic acid and peptide polymerization catalysts were set to 1 (*Coef1* and *Coef2*). Consequently, an increase in catalytic molecules leads to an accelerated elongation rate of polymers within the natural pool.

Furthermore, the efficiencies of polymerization, translation, and replication were globally suppressed according to the congestion level of the free-chain pools. Occupancy was defined as the total number of free nucleic acid and peptide chains divided by the sum of their respective pool capacities. The congestion coefficient was then determined as (1 – occupancy), with a lower limit of 0.05 to prevent complete cessation of reactions. This coefficient was applied as a multiplicative factor to all processes, including nucleic acid polymerization, peptide polymerization, translation, and RNA replication. This term serves as a coarse-grained representation of the reduced degrees of freedom for diffusion and molecular rearrangement that occurs as polymers accumulate within the finite volume of a protocell.

The translation reaction, which generates catalysts, was configured to occur only when both types of primordial tRNAs were present alongside the amino acid polymerization catalyst (c3) [Fig. 1(d)]. The targets for translation were not arbitrary sequences in the free pool, but specifically the protected "codon-recording nucleic acids." During each translation event, one record was randomly selected from the protected nucleic acid pool to produce its corresponding 9-bit catalytic peptide. Translation would fail if the required amino acid monomers were insufficient. The translation efficiency coefficient was set to 10 (*Coef3*) for the reference parameters. In this model, translation is implemented as the reproduction of functional peptides from protected nucleic acid records, triggered by the presence of tRNAs and polymerization catalysts.

Independently, when a new functional peptide was generated and exercising from the natural pool, it was transcribed into a protected RNA record as a sequence of codons, provided that both types of primordial tRNAs, the nucleotide polymerization catalysts (c0 and c1), and the writing auxiliary catalyst (c60) were available [Fig. 1(e)]. This process transcribes the functional peptide sequence into 3-bit codon strings on the nucleic acid side, consuming the necessary nucleic acid monomers. Thus, the establishment of a peptide function induces the formation of an RNA record, which subsequently serves as a template for future translation.

RNA replication was permitted only in the presence of the RNA replication catalyst (c61) [Fig. 1(f)]. The targets for replication included the two types of protected tRNAs and various codon-recording nucleic acids. During each replication event, one of these protected nucleic acids was randomly selected, and an additional copy was produced if sufficient nucleic acid monomers were present. The replication efficiency coefficient was set to 5 (*Coef4*). Replication in this model is specifically defined as the amplification of protected RNAs that have been crystallized as records, rather than a generalized replication of all free nucleic acid chains.

*2.5. Adjustment of Oscillation Centers and Information Feedback*

In the cognitive model, the centers of the two oscillators were sequentially adjusted by utilizing the bias of 0s and 1s within the group of protected functional molecules (refer to Appendix A.4 for details on the adjustment of oscillation centers). Specifically, the accumulated memory quantities corresponding to the "1" and "0" bits were aggregated based on the bit configurations of protected tRNAs, catalytic peptides, and codon-recording nucleic acids. In the cognitive model, these two quantities were compared every five steps. If the "1" side was dominant, the substrate amount of the long-period oscillator was decreased by 0.001, while that of the short-period oscillator was increased by 0.001. Conversely, if the "0" side was dominant, the substrate amount of the long-period oscillator was increased by 0.001, and that of the short-period oscillator was decreased by 0.001. In cases where the two quantities were equal, the direction of the adjustment was determined randomly. The variation range for each substrate amount was restricted between 0.1 and 5.0. Through this mechanism, the bit composition of functional molecules accumulated within the system was fed back into the 0/1 generation bias of the subsequent stage.

This feedback mechanism does not directly supply a target sequence from the external environment. Instead, the composition of functional molecules generated internally by chance influences the subsequent sequence generation process via the oscillation centers, thereby biasing the system’s overall exploration toward directions where functional molecules are more likely to be produced. Consequently, the process analogous to active inference in this model is implemented as an iteration of "exploration through fluctuations in internal states" and "readjustment of oscillation centers based on functional molecule composition."

*2.6. Entropy Measurement*

To quantify the bias in sequence exploration within the free pool, the Shannon entropy of 9-bit sub-sequences was calculated for each protocell. The analysis targeted both free nucleic acid and peptide chains, with entropy determined from the frequency distribution of all contiguous 9-bit sub-sequences contained within them. These calculations were performed every 10 steps. A reduction in entropy indicates that the distribution of 9-bit sequences in the free chain pool has deviated from a uniform random state and that specific sequence groups have become dominant—signifying a progressive narrowing of the sequence search space.

Notably, this metric is not intended to directly measure the absolute quantity of functional molecules retained in the protected pool; instead, it serves as an indicator of the progression of search bias within the free pool. Because the retention and inheritance of functional sequences are primarily mediated by the protected pool and its associated replication and translation processes, the decrease in entropy within the free pool was interpreted as an expression of "exploratory bias" rather than "total memory capacity."

*2.7. Evolution of the Protocell Population*

Generation turnover was implemented within the protocell population. Under reference conditions, the fitness of all protocells was calculated every 100 steps, and population selection was performed based on their rankings. The fitness $F$ of each protocell was defined as:

$$F = Fw_{c013} \times C_{poly} + Fw_{c60} \times C_{write/rep} + Fw_{tRNA} \times T + Fw_{codon} \times R + Fw_{trans} \times N_{tr} + Fw_{pool}\,(L_{NA,max} + L_{Pep,max})$$

where $C_{poly}$ is the total number of protected polymerization catalysts, $C_{write/rep}$ is the sum of sequence-writing auxiliary catalysts (c60) and nucleic acid replication catalysts (c61), $T$ is the total of the two types of primordial tRNAs, $R$ is the total number of various codon-recording nucleic acids, $N_{tr}$ is the count of successful translations in the preceding interval, and $L_{NA,max}$ and $L_{Pep,max}$ are the maximum lengths of free nucleic acid and peptide chains, respectively. The reference values for the fitness weights were set as: $Fw_{c013} = 10$, $Fw_{c60} = 20$, $Fw_{tRNA} = 15$, $Fw_{codon} = 5$, $Fw_{trans} = 5$, and $Fw_{pool} = 1$. This evaluation integrates polymerization functions, writing/replication capabilities, primordial translation, nucleic acid records, recent translation performance, and the extent of polymer elongation.

During generation turnover, the top half of the population by fitness was copied to replace the bottom half. This replication inherited the free pool, protected pool, monomer levels, and oscillator states, while introducing minute fluctuations only to the phases of the two oscillators. Consequently, generation turnover serves as an operation that simultaneously preserves functional configurations and introduces slight variations in oscillation phases. The translation count was reset after each evaluation to reflect the translation performance specific to each evaluation interval in the fitness score. Evolution in this model is thus described as dual dynamics, where functional molecule groups self-organized within internal reaction networks are selected and inherited at the population level.

### *2.8. Implementation of the Model*

The model was implemented using HTML and JavaScript. The software architecture consists of the following four programs, and the source code (in HTML format) is available via the URL provided in the "DATA AVAILABILITY" section. These programs can be executed in a standard web browser by saving the HTML file to a local computer. Various parameters can be configured through the user interface (UI), ensuring that all results presented in this paper are fully reproducible.

The four programs are:

1. **Model Simulation Program**: Conducts calculations for a single random seed.
2. **Visualization Program (Single Result)**: Displays results from a single random seed calculation.
3. **Batch Simulation Program**: Executes sequential calculations across 30 consecutive random seeds.
4. **Visualization Program (Aggregate Results)**: Displays the average results from the batch calculations.

For random number generation, the Xorshift algorithm [38] was implemented. Proposed by George Marsaglia in 2003, Xorshift is a pseudo-random number generation algorithm characterized by its high execution speed and simplicity of implementation.

## 3. Results

### *3.1. Simulation Results of the Reference Case*

Table 1 presents the parameter set for the reference case. Due to the large number of parameters in this model, the parameter space is vast; however, the extent of the regions yielding significant results remains unclear. Preliminary calculations were performed to identify a parameter set that produces meaningful outcomes, which was then established as the reference for evaluating the model's superiority. Calculations for both the cognitive and random models were conducted using a base random seed of 20.

**Table 1.** Reference parameters.

| | | |
|---|---|---|
| Basic settings | Base Random Seed | 20、40、60、80、100 |
| | Max Time Steps | 2000 |
| | Model Type | Cognitive, Random |
| | Calculation time step; dt | 0.08 |
| | Number of protocells | 50 |
| | Protocell evolution interval | 100 |
| Setting of polymer polymerization and decomposition | Initial natural polymerization rate | 0.04 |
| | Initial natural decomposition rate | 0.14 |
| | Increase in natural decomposition rate (per step) | 0.000002 |
| | Mutation rate | 0.001 |
| Initial number of monomers | Initial number of amino acids | 50000 |
| | Initial number of nucleotides | 50000 |
| Parameters of oscillator A | Initial amount of $A_A$ | 1 |
| | Initial amount of $X_A$ | 9 |
| | Initial amount of $Y_A$ | 10 |
| | Initial amount of $k_{1A}$ | 0.8 |
| | Initial amount of $k_{2A}$ | 0.1 |
| | Initial amount of $k_{3A}$ | 0.6 |
| Parameters of oscillator A | Initial amount of $A_B$ | 0.5 |
| | Initial amount of $X_B$ | 9 |
| | Initial amount of $Y_B$ | 10 |
| | Initial amount of $k_{1B}$ | 1.5 |
| | Initial amount of $k_{2B}$ | 0.1 |
| | Initial amount of $k_{3B}$ | 0.9 |
| Threshold of composite wave | Threshold θ | -6 |
| Target array | tRNA-1 | 101110101 |
| | tRNA-2 | 101110111 |
| | Catalyst-c0 | 111010111 |
| | Catalyst-c1 | 111001111 |
| | Catalyst-c3 | 111101111 |
| | Catalyst-c60 | 101010101 |
| | Catalyst-c61 | 101101101 |
| Humming distance | Coefficient of Hamming distance 1 (Catalyst) | 0.1 |
| | Coefficient of Hamming distance 2 (Catalyst) | 0.05 |
| | Coefficient of Hamming distance 1 (tRNA) | 0.01 |
| Coefficient of catalytic reaction | Translation efficiency coefficient; Coef1 | 1 |
| | Translation efficiency coefficient; Coef2 | 1 |
| | Translation efficiency coefficient; Coef3 | 10 |
| | Translation efficiency coefficient; Coef4 | 5 |
| Parameters of the vibration center | Change in the center of vibration per step | 0.001 |
| Coefficients in the fitness calculation formula | Fitness weight; Fw_c60 | 20 |
| | Fitness weight; Fw_tRNA | 15 |
| | Fitness weight; Fw_c013 | 10 |
| | Fitness weight; Fw_codon | 5 |
| | Fitness weight; Fw_trans | 5 |
| | Fitness weight; Fw_pool | 1 |

Figure 2 illustrates the macroscopic analysis results for the cognitive and random models in the reference case. These plots represent the temporal transitions of the average state variables across 50 protocells. Regarding the entropy transition (Figure 2A), entropy in the cognitive model begins to decrease sharply around step 300, synchronized with an exponential increase in functional molecules (Figure 2B). This observation provides evidence that the "learning of meaningful patterns" by the oscillators is functioning as intended. An examination of the stacked graph in Figure 2B from the base upward reveals that the tRNA layer (orange) thickens first, subsequently driving the emergence of other catalysts. This process visualizes the dynamics where primordial tRNAs, bridging amino acids and nucleic acids, emerge first and serve as a scaffold for the subsequent emergence of translation and replication systems. In contrast, the entropy in the random model remains high (Figure 2A), and the number of functional molecules remains near zero (not shown in Figure 2). These results demonstrate that the cognitive model generates catalytic peptides and RNA significantly faster and more frequently than the random model.

The transitions of maximum polymer lengths (Figures 2C and 2D) show that lengths in the cognitive model converge toward the maximum of 100, whereas the random model fails to generate long molecules. Figures 2E and 2F show the number of protocells where catalytic reaction loops (translation and replication systems) have been established. In Figure 2E, the curve rises with exponential increase. This captures the moment the "autocatalytic amplification loop" is ignited as tRNAs, translation catalysts, and replication catalysts align, rather than molecules being produced in isolation.

From these results, the following emergent behaviors are inferred to occur within the cognitive model:

1. **Random Exploration**: Initially, meaningless sequences are produced through random polymerization and decomposition.
2. **Accidental Discovery**: Sequences possessing tRNA or catalytic functions are generated by chance when polymerization coincides with the patterns produced by the synthesis of Oscillators A and B.
3. **Local Amplification**: Catalysts facilitate the transcription of peptide sequences into RNA, as well as tRNA polymerization and replication.
4. **Emergence of Catalytic Loops**: As catalytic loops assisting in peptide production and RNA polymerization/replication begin to operate, the counts of tRNAs and catalysts increase rapidly. Consequently, the protocell becomes populated with specific functional sequences.

Results regarding the influence of other parameters, excluding those presented in Sections 3.1 to 3.3, are provided in "Appendix F: Sensitivity Analysis of Parameters."

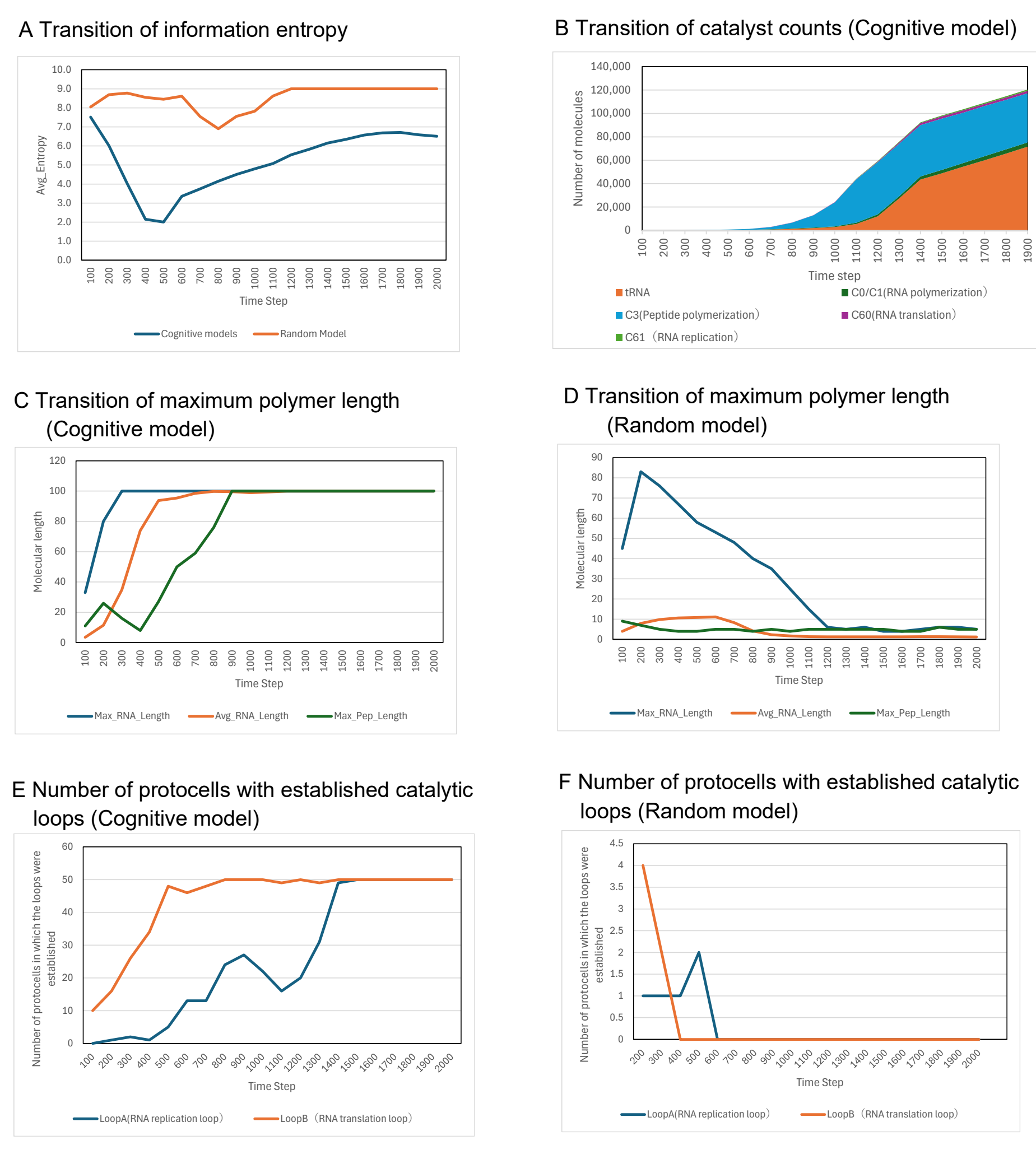

**Figure 2. Macroscopic analysis results for the reference parameter case. Comparison between the cognitive and random models. (A)** Transition of information entropy; the decrease in the cognitive model indicates the emergence of a meaningful bias in the polymer pool. **(B)** Transition of catalyst counts in the cognitive model; tRNAs and catalytic sequences increase as entropy decreases. **(C, D)** Comparison of maximum polymer length; length-100 molecules become dominant in the cognitive model, while the random model fails to produce long molecules. **(E, F)** Number of protocells with established catalytic loops; the number increases in the cognitive model but remains at zero in the random model.

Figure 3 shows the transitions of internal variables for the protocell with the highest fitness in the reference case (cognitive model). This figure illustrates the temporal evolution of how the chemical oscillators perform "learning" to generate functional molecules. Focusing on the "Transitions of *Mplus*, *Mzero* and migration of oscillation centers" in Figure 3C, it is observed that the substrate amounts (A, B) are dynamically adjusted in the early stages based on the difference between *Mplus* and *Mzero*. This reflects the system searching for the "optimal waveform" most likely to generate target sequences from an undifferentiated state.

Around step 800, *Mplus* and *Mzero* converge to a constant ratio, synchronized with an exponential burst in the molecular counts of tRNAs and various catalysts. This transition captures the "phase transition" where the interference patterns of multiple oscillators align with the target sequences, establishing the autocatalytic loop.

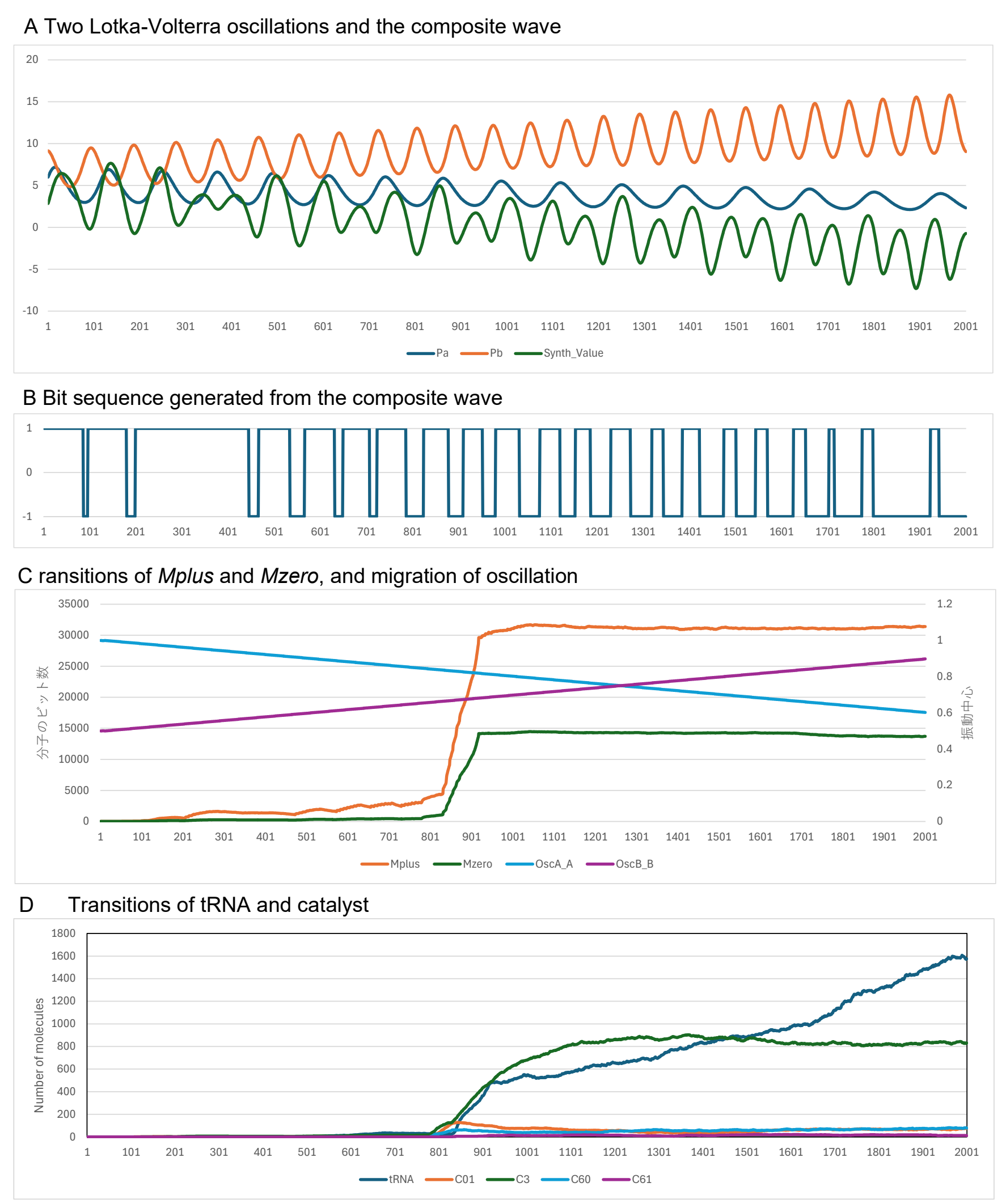

**Figure 3. Internal variable transitions for the protocell with the highest fitness in the reference case (cognitive model). (A)** Waveforms of the two Lotka-Volterra oscillators and their composite wave (blue). **(B)** Bit sequence waveform generated from the composite wave using the threshold. **(C)** Transitions of *Mplus* and *Mzero* calculated from the totals of 0s and 1s in the polymer pool, showing the resulting migration of oscillation centers. **(D)** Transitions of tRNA and catalyst counts; a rapid increase is observed following the establishment of the catalytic reaction loop.

### *3.2. Investigation of the Success Rate of Catalytic Loop Emergence*

To evaluate the validity of the model across various random seeds, a batch processing program was utilized to perform calculations for 30 consecutive random seeds under the reference parameter set. In this batch analysis, the success rate of catalytic loop emergence was evaluated based on the following criteria:

**Criteria for Establishing a Catalytic Loop**: A catalytic loop is considered to have emerged if at least one protocell appears within the total simulation duration (e.g., 2000 steps) where all seven functional molecules—tRNA-1 (polymerization of amino acid 0), tRNA-2 (polymerization of amino acid 1), c0 (RNA polymerization 0), c1 (RNA polymerization 1), c3 (peptide polymerization), c60 (reverse transcription/separation), and c61 (RNA replication)—are present for 10 or more time steps.

Table 2 summarizes the success rates of catalytic reaction loops. For example, "Random Seed 20" indicates the aggregate success rate for 30 calculations performed with random seed values ranging from 5 (= 20 - 15) to 34 (= 20 + 14). The results indicate that the success rate in the cognitive model is significantly higher than that of the random model across all random seeds. This highlights the effectiveness of the oscillator-driven induction of polymerization. The high success rate of approximately 80% in the cognitive model suggests that the self-organized crystallization of information from initial random fluctuations can occur with extremely high probability. This supports the possibility that the origin of life was not a miraculous coincidence but a physicochemical necessity under non-equilibrium driving.

The success rate in the random model remained low but non-zero. This is attributed to the fact that the defined catalysts and tRNAs are only 9-bit binary patterns and that catalytic activity based on Hamming distance is permitted, allowing catalysts to emerge with a certain probability even through random polymerization.

**Table 1.** Success Rate of Catalytic Loop Emergence.

| Random Seed Group | 20 | 40 | 60 | 80 | 100 |
|---|---|---|---|---|---|
| Cognitive Model | 83.3 % | 83.3 % | 80.0 % | 76.7 % | 83.3 % |
| Random Model | 23.3 % | 20.0 % | 13.3 % | 20.0 % | 10.0 % |

Note: For each "Random Seed X," results are aggregated from 30 calculations using seeds ranging from X - 15 to X + 14.

*3.3. Investigation of Success Rates with Different Target Sequences*

The possibility of catalytic loop establishment was examined when the target sequences were modified. Table 3 defines the alternative target configurations, and the success rates for these cases using Random Seed 20 (seeds 5 to 34) are summarized in Table 4.

The results show that the success rate of the catalytic loop remains significantly higher in the cognitive model than in the random model, even when target sequences are changed. This demonstrates the "generalization" capability of the model. It suggests that the cognitive model can self-adjust its internal waveforms to form a loop not only for one specific sequence but also for sequences with different ratios of 0s and 1s. This implies a versatile architecture capable of adapting to unknown environments rather than being limited to a specific "correct" solution.

The fact that success rates are maintained across different target sequences (Table 4) indicates that the model is not dependent on a specific set of target strings. Instead, it represents a generalized architecture that can self-adjust its rhythms to extract "meaning" (functional sequences) in response to varying environmental demands (targets).

**Table 3.** Configuration of Target Sequences.

| Target Molecule | Case 1 (Ref) | Case 2 | Case 3 | Case 4 |
|---|---|---|---|---|
| **tRNA-1** | 101110101 | 101010101 | 111110111 | 100110001 |
| **tRNA-2** | 101110111 | 101101101 | 111111111 | 100110011 |
| **Catalyst-c0** | 111010111 | 111010111 | 111010111 | 110010011 |
| **Catalyst-c1** | 111001111 | 111001111 | 111001111 | 110001011 |
| **Catalyst-c3** | 111101111 | 111101101 | 111101111 | 110101011 |
| **Catalyst-c60** | 101010101 | 101110101 | 101010101 | 100010001 |

| Catalyst-c61 | 101101101 | 101110111 | 101101101 | 100101001 |
|---|---|---|---|---|

**Table 4.** Success Rate of Catalytic Loop Emergence with Different Target Sequences (Random Seed 20 [5–34]).

| Target Case | Case 1 (Ref) | Case 2 | Case 3 | Case 4 |
|---|---|---|---|---|
| Cognitive Model | 83.3 % | 53.3 % | 76.7 % | 53.3 % |
| Random Model | 23.3 % | 10.0 % | 10.0 % | 26.7 % |

# 4. Discussion

## *4.1. Key Findings: Internal Oscillations as an Effective Bias for Sequence Search*

The primary result of this study is that the cognitive model, utilizing two chemical oscillators, achieved a higher frequency and greater stability in forming catalytic loops compared to the random control model. These loops consist of functional peptides, primordial tRNAs, and RNA molecules responsible for their recording and amplification. Specifically, the fact that the cognitive model maintained high success rates across all random seeds in the reference case, while those of the random model significantly declined, suggests that the bias in 0/1 selection driven by internal oscillations facilitates the exploration of sequence regions difficult to reach through purely random polymerization. This implies that the emergence of functional molecules can be accelerated by effectively narrowing the search space through temporal bias dependent on internal states, rather than through uniform exploration of the entire sequence space.

Furthermore, this model does not merely facilitate the generation of specific sequences through chemical oscillators; it incorporates a closed loop where generated functional molecules are transferred to a "protected functional molecule pool" for subsequent reuse in translation, nucleic acid recording, and replication. This functional molecule pool represents naturally occurring phenomena such as the folding of amino acid chains, where reduced degradability arises from such folding. Consequently, the system adopts a two-stage structure where internal oscillations function at the entry point of sequence generation, while the mechanisms for protection via folding, recording, and re-amplification govern the accumulation of functional molecules. This architecture ensures that accidentally generated functional sequences are not immediately lost to natural degradation but persist as a historical record within the system, influencing reactions in subsequent generations. Thus, this study is characterized by modeling chemical oscillations not merely as periodic phenomena, but as search engines equipped with information retention capabilities.

## *4.2. Implications for the Emergence of the "Catalyst-RNA System"*

A central challenge in the origin of life is understanding how catalytic peptides and information-bearing nucleic acids emerged as an interdependent system. The results of this study provide computational evidence for a scenario where bridging molecules equivalent to primordial tRNAs appear early and serve as a scaffold for recording and reproducing catalytic functions.

This aligns with the "metabolism-first" perspective proposed by E. Smith and H. J. Morowitz [39], which posits that life originated from the rhythms of cyclic metabolic networks as an inevitable consequence of geochemical energy flows. The concept of a "chord of chemical oscillations" in this model serves as a computational working hypothesis explaining how the physical rhythms of the prebiotic environment were "frozen" into digital records of polymer sequence information.

In the results, a trend was observed in the cognitive model where primordial tRNA groups emerged first, followed by an increase in various catalytic peptides and codon-recording nucleic acids. This suggests that the translational intermediate structures linking amino acids and nucleic acids may be one of the rate-limiting steps in the establishment of catalytic loops. While the GADV hypothesis [7-8] assumes a stepwise progression from the appearance of weak globular peptide catalysts to the subsequent establishment of primordial tRNAs, this model demonstrates that this transition can be understood as a result of search biased by internal oscillations rather than purely random exploration.

Concurrently, this study is not intended to replace the RNA world hypothesis or the peptide-RNA co-evolution theory. Instead, it is positioned as a hypothetical model that adds exploratory bias from internal oscillations and mechanisms for recording and protection to the conditions for self-sustaining reaction sets described by autocatalytic networks [24-26] and RAF theory [27]. For an autocatalytic network to be established, the sequences and molecular species constituting the network must appear with appropriate probabilities, rather than merely being linked. These findings suggest that chemical oscillations can function as a mechanism to provide this biased probability of appearance.

### *4.3. Relationship with Previous Research: A Framework Linking Compartmentalization, Autocatalysis, and Periodic Driving*

In previous research, liquid-liquid phase separation (LLPS) and coacervates have attracted attention as sites responsible for molecular concentration, memory, and responsiveness [22-23]. Additionally, autocatalytic networks and Reflexively Autocatalytic and Food-generated (RAF) theory have mathematically defined the self-maintenance of reaction sets [24-27]. Furthermore, studies on Belousov-Zhabotinsky (BZ) reactions and supramolecular oscillators have experimentally demonstrated that periodic phenomena can control polymerization and structural formation [29-33]. The significance of this study lies not in the direct reproduction of these individual elements, but in providing an integrated perspective where periodic driving biases sequence exploration within compartmentalized internal spaces, resulting in the formation of autocatalytic and recording molecular networks. Thus, this model can be viewed as an abstract framework that introduces chemical oscillations as an element providing temporal order to the emergence of autocatalytic networks within LLPS-like compartments.

Moreover, compared to the discussion that external cycles such as wet-dry cycles can promote prebiotic polymerization [34-35], this model can be interpreted as considering the "internalization" of such periodic driving within the cell. If periodic fluctuations in the external environment were incorporated as internal chemical oscillations during evolution, the processes of polymerization, selection, and recording would no longer depend passively on external conditions but would become self-adjusting according to internal states. The fact that the cognitive model demonstrated a higher success rate than the random model in this study supports the possibility that the transition from external to internal cycles favors the emergence of informational molecules.

Furthermore, a comparison with the Graded Autocatalysis Replication Domain (GARD) model [36] clarifies the positioning of this work. While GARD deals with composition-based information based on compartmentalization and mutual catalysis, this study addresses sequence-based informational molecules through 0/1 sequences. Therefore, while GARD focuses its discussion on compositional information, namely, what is present and in what quantities, this model focuses on the establishment of sequence information, the order in which units are arranged. In this sense, this model is positioned as a computational framework for considering the bridge from pre-genetic compositional information to genetic sequence information.

*4.4. Significance of Success Rates and Sensitivity Analysis: The Model is Robust but Not Unconditionally Valid*

In the batch calculations conducted across various random seeds, the cognitive model consistently demonstrated higher success rates for catalytic loop emergence than the random model. Furthermore, this superiority was largely maintained even when the target sequences were modified. These findings indicate that the model is not an "artificial construct" excessively dependent on a single specific sequence; rather, the exploratory bias driven by internal oscillations and the mechanisms for protection, recording, and amplification function with a degree of generality. Therefore, the essence of the model lies in the dynamic principles that bias sequence exploration, rather than in the hard-coding of specific target strings.

On the other hand, the sensitivity analysis provided in Appendix F demonstrates that this superiority is not unconditional. The significant variation in success rates following changes in the calculation time step ($dt$) implies that the sampling interval for reading 0/1 patterns from the oscillators is critical. In this model, the time step functions not merely as a numerical setting but as a physiological and informational parameter determining the timing for mapping internal oscillations onto polymer sequences. Furthermore, since altering the Hamming distance weights for catalysts and tRNAs caused substantial fluctuations in success rates, the degree to which weak functional sequences are tolerated determines the initiation of the early reaction network. Additionally, variations in success rates observed with changes in the number of protocells and fitness weights indicate that both internal reaction dynamics and the design of group selection influence the ease of emergence. Consequently, while the model operates within a relatively broad range of conditions, it does not perform identically under all conditions. This behavior is natural for a model of the origin of life and serves as a clue for identifying the specific conditions required for emergence.

*4.5. Significance of Information Entropy Reduction: Transition to the Subject of Information Generation*

In this study, information localization within the system was tracked using the Shannon entropy of sequence distribution. The fact that entropy decreased while functional molecule groups increased in the cognitive model reflects a process where specific 9-bit sequences become dominant within the protocell, rather than a mere increase in molecular count. In other words, the system moves away from a uniform random sequence distribution and toward the selective accumulation of specific sequences dependent on internal states. In this regard, the model is understood not as a "replication system provided with a correct sequence from the outside," but as a system that biases its own sequence distribution through internal reaction history and the readjustment of oscillators.

However, caution is required before asserting that this entropy reduction directly represents the acquisition of "meaning" or "knowledge." This research measures information localization based solely on the frequency distribution of sub-sequences, which is not equivalent to semantic information or external representation itself. Therefore, it is more appropriate to interpret these results as demonstrating that the internal reaction network forms a structure that favors specific sequences and that this structure is reinforced over time. Furthermore, from the perspective of information thermodynamics [37], such a decrease in entropy can be positioned as a coarse-grained indicator of memory formation in non-equilibrium systems accompanied by dissipation. In short, it is more reasonable to state that this study demonstrates the formation of localization and recording, the physical foundations of information, rather than the completed form of semantic information.

*4.6. Limitations of the Model*

Several significant limitations exist in this model. First, the nucleic acid and peptide chains dealt with in this study are binary (0/1) sequences, which do not directly represent the four types of actual nucleotides or twenty types of amino acids. Consequently, this research is not a realistic reconstruction of the origin of life but rather an abstract model extracting the logical structures of combinatorial explosion, functional sequence search, and the closed loop of recording and amplification. Second, functional molecules are treated as predefined target sequences; the model has not yet reached the stage of evolutionarily determining which sequences possess function. Third, the protected pool, translation, nucleic acid recording, and replication are all represented by simplified rules, omitting factors such as three-dimensional structural formation, binding/dissociation equilibrium, energy currencies, and the spatial heterogeneity of reaction sites.

Furthermore, the mapping of temporal patterns onto spatial sequences remains at an abstract stage in this study. While the mechanism by which the timing of internal oscillations is converted into 0/1 additions at polymer ends is conceptually clear, it remains unresolved which specific reaction elements, interfaces, metal catalysts, or phase-separated structures facilitate this process in actual chemical systems. These points require further concretization through connections with experimental studies on BZ-driven polymerization and supramolecular oscillators [29-33], polymerization experiments via wet-dry cycles [34,35], or compartmentalized systems [22,23].

*4.7. Future Perspectives*

The primary challenge for future research is to expand the model by increasing the variety of monomers. While this study utilized binary sequences, future extensions should incorporate four types of nucleotides and four or more types of amino acids to evaluate the effectiveness of the proposed exploratory bias and recording/amplification mechanisms within a more realistic sequence space.

Second, the expansion from two chemical oscillators to a multi-oscillator system is necessary to investigate whether richer sequence patterns can be generated through combinations of amplitude, period, phase, and threshold. As shown in Appendix E, the synthesis of only two waves or two Lotka-Volterra oscillations can generate diverse 0/1 patterns; this direction offers a direct path to verify the central hypothesis of this study, the "chord of chemical oscillations."

Third, a bridge between the abstract model and experimental systems must be established. For instance, the influence of internal cycles on polymerization bias and polymer length distribution should be verified using compartmentalized coacervates or BZ-driven polymerization systems [34,35,29-33].

Fourth, since success rates are affected by fitness functions and population sizes, a systematic investigation of selection pressures at various evolutionary stages is required. The system's state can vary depending on whether priority is given to translation, replication, recording, or elongation. This underscores the necessity of treating the origin of life as a search problem with multiple local optima rather than a single global solution. Future efforts should integrate internal reaction networks, information localization, and group selection into a unified framework to identify which prebiotic architectures are favored under specific conditions.

## 5. Conclusions

In this study, a cognitive model was developed that imposes an intrinsic bias on binary (0/1) selection during polymer elongation through the use of two chemical oscillators. The investigation focused on the emergence of catalytic loops composed of functional peptides, primordial tRNAs, and the RNA molecules that record and amplify them. The results demonstrated that the proposed model exhibits superior performance compared to a random control model in terms of the establishment rate of catalytic loops, the accumulation of functional molecules, the elongation of polymers, and the reduction of entropy in sequence distribution. These findings suggest that internal chemical oscillations effectively narrow the sequence search space and facilitate the formation of functional molecular assemblies.

While this model is not intended to be a direct reproduction of the origin of life, it provides a computational working hypothesis for understanding the process by which catalytic function and information retention emerge interdependently. Future research should verify the generality and empirical validity of this framework by expanding the types of monomers, developing multi-oscillator systems, and establishing correspondences with experimental compartmentalized systems.

**Supplementary Materials:** There is no Supplementary Materials.

**Author Contributions:** T.I. conceptualized the methodology, software development, formal analysis, and investigation of the study. T.I. also drafted and revised the manuscript, gave final approval for publication, and agreed to be accountable for all aspects of the work.

**Funding:** This research was supported by grants from the Japan Society for the Promotion of Science, KAKENHI Grant Number 23K04283.

**Data Availability Statement:** All data supporting the findings of this study are available within the main text and appendix. The source code (HTML + JavaScript) used for the simulations is available on GitHub. The results presented in this paper can be reproduced by changing the UI parameters of the program.

https://github.com/Takeshi-Ishida/Chord-of-Chemical-Oscillations-Emergence-of-Catalyst-RNA-Systems

A preprint of this manuscript has previously been published on arXiv:XXXXXX.

**Acknowledgments:** During the preparation of this manuscript, the author(s) used Gemini 3 Pro and Chat GPT 5.4 for the purposes of brainstorming research ideas, generating program code, and performing English translation of the text. The authors have reviewed and edited the output and take full responsibility for the content of this publication.

**Conflicts of Interest:** The author declares that there are no conflicts of interest regarding the publication of this paper. The research was supported by the funding mentioned in the Finding; however, the funder had no role in the study design, data collection, analysis, interpretation of data, or the writing of the manuscript.

# Appendix A

*Appendix A.1 The Lotka-Volterra Chemical Oscillation Model*

A.1.1 Overview of the Standard Form of the Lotka-Volterra Equations

The standard form of the Lotka-Volterra equations, representing the most classic predator-prey model, is a system of nonlinear differential equations that describes the change over time in the population (or concentration) of two biological (or chemical) species.

**$X$**: The population or concentration of the prey (or a self-replicating chemical species).

**$Y$**: The population or concentration of the predator (or a chemical species that consumes X to replicate).

The equations are expressed as follows:

$$dX/dt = X(a - bY)$$
$$dY/dt = Y(cX - d)$$

Here, the parameters have the following meanings:

**$a$**: Intrinsic natural growth rate of prey $X$ (related to the growth rate of $X$ in the absence of predator $Y$; $a > 0$).

**$b$**: Efficiency of predation (consumption) of prey $X$ by predator $Y$ ($b > 0$).

**$c$**: Efficiency of the increase in predator $Y$ from preying on (consuming) prey $X$ ($c > 0$).

**$d$**: Intrinsic natural decay rate of predator Y (decay rate of Y in the absence of prey X; $d > 0$).

This model was originally proposed to explain the periodic population fluctuations of predators and prey in ecosystems, but it can also be applied to chemical reaction systems that exhibit similar dynamics, such as networks of autocatalytic or enzymatic reactions. The most notable feature of this model is that when the parameters are within an appropriate range, the populations (concentrations) of $X$ and $Y$ continue to oscillate periodically.

A.1.2 Describing the Standard Form of the Lotka-Volterra Equations with Chemical Reactions

The chemical reaction scheme corresponding to the above system of equations is as follows:

**1. Self-replication of substance (*X*)**:

$A \rightarrow X$ (with rate constant $k_1$)

$X \rightarrow 2X$ (with rate constant $a = k_1 A$)

(This represents the conversion of an input substance $A$ into $X$, and the self-replication of $X$).

**2. Consumption of substance (*X*) by substance (*Y*) and the resulting replication of substance (*Y*):**

$X + Y \rightarrow 2Y$ (with rate constant $b = k_2$)

(This reaction, where $X$ and $Y$ react, consumes one molecule of $X$ and results in a net increase of one molecule of $Y$, contributes to both the decay term for X, $-bXY$, and the growth term for $Y$, e.g., $bXY$, depending on the setup of parameter $c$).

**3. Self-decay of the predator (*Y*):**

$Y \rightarrow P$ (where $P$ is an inert product or removed from the system) (with rate constant $d = k_3$)

(This represents the decomposition or inactivation of one molecule of $Y$).

By formulating differential equations based on the law of mass action from these reactions, the standard form is obtained. (In correspondence with the standard system's equations, $a=k_1A$, $b=k_2$, $c=k_2$, and $d=k_3$. The program implementation uses $A$, $k_1$, $k_2$, and $k_3$ as parameters) .

A.1.3 Period and Amplitude of Lotka-Volterra Oscillations

**(1) Control of Oscillation Period**

The oscillation period $T$ is primarily determined by the product of parameters $a$ (prey growth rate) and $d$ (predator decay rate). In linear approximation (small oscillations near the equilibrium point), the period is approximated by the following formula:

$$T = \frac{2\pi}{\sqrt{ad}} = \frac{2\pi}{\sqrt{k_1 A k_3}}$$

Methods to shorten the period (increase oscillation speed):

- Increase $A$ or $k_1$: Increasing the substrate supply $A$ or the reaction rate $k_1$ raises the intrinsic growth rate a of the prey $X$, thereby accelerating the cycle.
- Increase $k_3$: Raising the decay rate d of the predator $Y$ accelerates the population adjustment cycle.

Methods to lengthen the period (decrease oscillation speed):

- Set lower values for the parameters $A$, $k_1$, or $k_3$.

In this model, the distinction between "Oscillator A (long period)" and "Oscillator B (short period)" is generated by adjusting the product of $k_1$, $A$, and $k_3$.

**(2) Control of Amplitude**

In the Lotka-Volterra model, the amplitude depends more strongly on the distance between the initial values and the fixed point than on the parameters themselves.

- The closer the initial values ($X_0$, $Y_0$) are to the fixed point ($X_s$, $Y_s$), the smaller the amplitude becomes.
- The further the initial values are from the fixed point, the larger the amplitude becomes.

**(3) Dependency on Initial Values**

The standard Lotka-Volterra equations adopted in this model do not include "dissipation" (energy loss) or "saturation," and thus possess the following characteristics: (a) Existence of conserved quantities: The system possesses a type of "energy" (conserved quantity), and the trajectory is entirely determined by the initial values. (b) Lack of convergence: Unless subjected to external perturbations, the system does not converge to a specific amplitude (limit cycle) over time. It continues to revolve around the fixed point in a neutrally stable orbit determined by the initial values. (c) Note on simulation: Moving the production amount $P$ through the "exploration mode (perturbation)" or "learning rules" mentioned in the paper is mathematically equivalent to moving the center of oscillation (fixed point). This relatively moves the current state further from or closer to the fixed point, consequently changing the amplitude.

A.1.4 Describing the Standard Form of the Lotka-Volterra Equations with Chemical Reactions

The center of oscillation for $X$ and $Y$ in the standard Lotka-Volterra equations is the system's non-trivial fixed point (equilibrium point). This is the point where $dX / dt = 0$ and $dY / dt = 0$ are simultaneously satisfied, and where $X \neq 0$ and $Y \neq 0$.

From the equations $X(a - bY) = 0$ and $Y(cX - d) = 0$, for the case where $X \neq 0$ and $Y \neq 0$, we get:

$a - bY = 0 \Longrightarrow Y_s = a / b$, and $cX - d = 0 \Longrightarrow X_s = d / c$.

Therefore, the center of oscillation (the non-trivial fixed point) is $(X_s, Y_s) = (d/c, a/b)$. According to the program's parameter settings ($a = k_1A$, $b = k_2$, $c = k_2$, $d = k_3$):

$X_s = k_3 / k_2$

$Y_s = (k_1A) / k_2$

This shows that $Y_s$ can be controlled by the magnitude of $A$. Since the production rate of $P$ is $k_3Y$, the oscillation center of $P$'s production can be controlled by $A$ or $k_1$. This fixed point is a "center," and the trajectories around it are neutrally stable, meaning they neither spiral into nor are repelled from the fixed point, but rather continue to oscillate periodically along a closed orbit determined by the initial conditions.

A.1.5 Construction and Solution of the Numerical Model

This model uses the 4th-order Runge-Kutta method (RK4). This method has the advantage of higher accuracy and yields stable solutions even with a relatively large time step (dt) compared to simpler methods like the Euler method.

## Appendix B
## Calculation Procedure for Information Entropy in This Model

Shannon entropy is utilized as a quantitative indicator to demonstrate that information generation can occur through internal processes alone. The "reduction in polymer sequence diversity (i.e., order formation)" within the protocell is quantified as the "entropy of sequence distribution." Specifically, the calculation is performed using the frequency distribution of 9-bit sequences in the natural pool, rather than analyzing all molecules within the cell.

**Specific Calculation Procedure**

The calculation is performed at each time step (or at fixed intervals) of the simulation according to the following steps:

**1. Definition of Analysis Object (Definition of Information)**

Since catalytic functions are defined based on specific 9-bit sequences, the analysis targets sub-sequences of length $L$ (k-mers).

- Setting: $L = 9$ (matching the length of the target sequences)
- Data Collection: All contiguous 9-bit sub-sequences are extracted and collected from every RNA and peptide chain present in the natural polymerization/decomposition pool within the protocell.

**2. Creation of Distribution**

The frequency of appearance for each collected sequence pattern is counted.

- Total number of patterns (denominator): $N$
- Occurrence count of a specific sequence pattern $i$ (e.g., 111010111): $n_i$
- Probability of appearance for that pattern: $p_i = n_i / N$

**3. Shannon Entropy Formula Entropy H(t) is calculated using the following equation:**

$$H(t) = -\sum_{i} p_i \; log_2 p_i$$

- Units: bits
- Note: $\sum_i \; p_i \; log_2 p_i$) represents the summation over all unique sequence patterns that appeared.

## Appendix C
## Configuration of Weak Activity Based on Hamming Distance of Target Sequences

In this model, sequences with a Hamming distance of 1 or 2 from the target sequence are configured to possess weak activity (activity coefficient < 1).

**1. Definition of Hamming Distance**

Hamming distance represents the number of positions at which the corresponding symbols are different between two strings of equal length. In the context of this binary simulation model (0/1 sequences), it serves as a metric for the "degree of dissimilarity," indicating how many bits (0 or 1) must be flipped to match the target sequence.

**Specific Examples (for a 9-bit sequence)**

Using target sequence 111010111 as an example:

- Case A: Exact match (Distance 0)
  Target: 1 1 1 0 1 0 1 1 1
  Query: 1 1 1 0 1 0 1 1 1
  No differences. Hamming distance = 0 (possesses strong activity).

- Case B: One-bit difference (Distance 1)
  Target: 1 1 1 0 1 0 1 1 1
  Query: 1 1 1 0 0 0 1 1 1
  The 5th bit differs. Hamming distance = 1 (assigned weak activity).

- Case C: Two-bit difference (Distance 2)
  Target: 1 1 1 0 1 0 1 1 1
  Query: 1 1 1 1 1 1 1 1 1
  The 4th and 6th bits differ. Hamming distance = 2 (assigned weak activity).

**2. Significance of Setting Weak Activity (Evolutionary Scaffolding)**

The evaluation based on Hamming distance is critical from the perspective of "evolutionary scaffolding" described in the paper.

- All-or-Nothing Scenario: If only exact matches (Distance 0) are permitted, the probability of coincidentally matching all 9 bits through random exploration is $1 / 2^9 = 1 / 512$. In the early stages, such events rarely occur, and evolution fails to initiate.
- Gradation Scenario (Considering Hamming Distance): Allowing a distance of up to 2 dramatically increases the number of acceptable patterns (utilizing the inverse of combinatorial explosion).

Distance 0: 1 pattern
Distance 1: 9 patterns (any one of the 9 positions differs)
Distance 2: 36 patterns (${}_9C_2$)

Total: 46 patterns

This allows the system to "climb the evolutionary stairs" by first generating molecules with weak activity (Distance 2), which then increase and mutate (Distance 1), eventually reaching full catalytic activity (Distance 0).

## Appendix D
## Examples of Binary (0/1) Patterns Generated from the Superposition of Two Sine Waves

This section provides examples of binary (0/1) patterns generated by applying a constant threshold to the combined waveform of two sine waves. The red lines in the graphs represent the 0/1 patterns. It is observed that a diverse range of patterns can be produced by varying the amplitude and the center of oscillation.

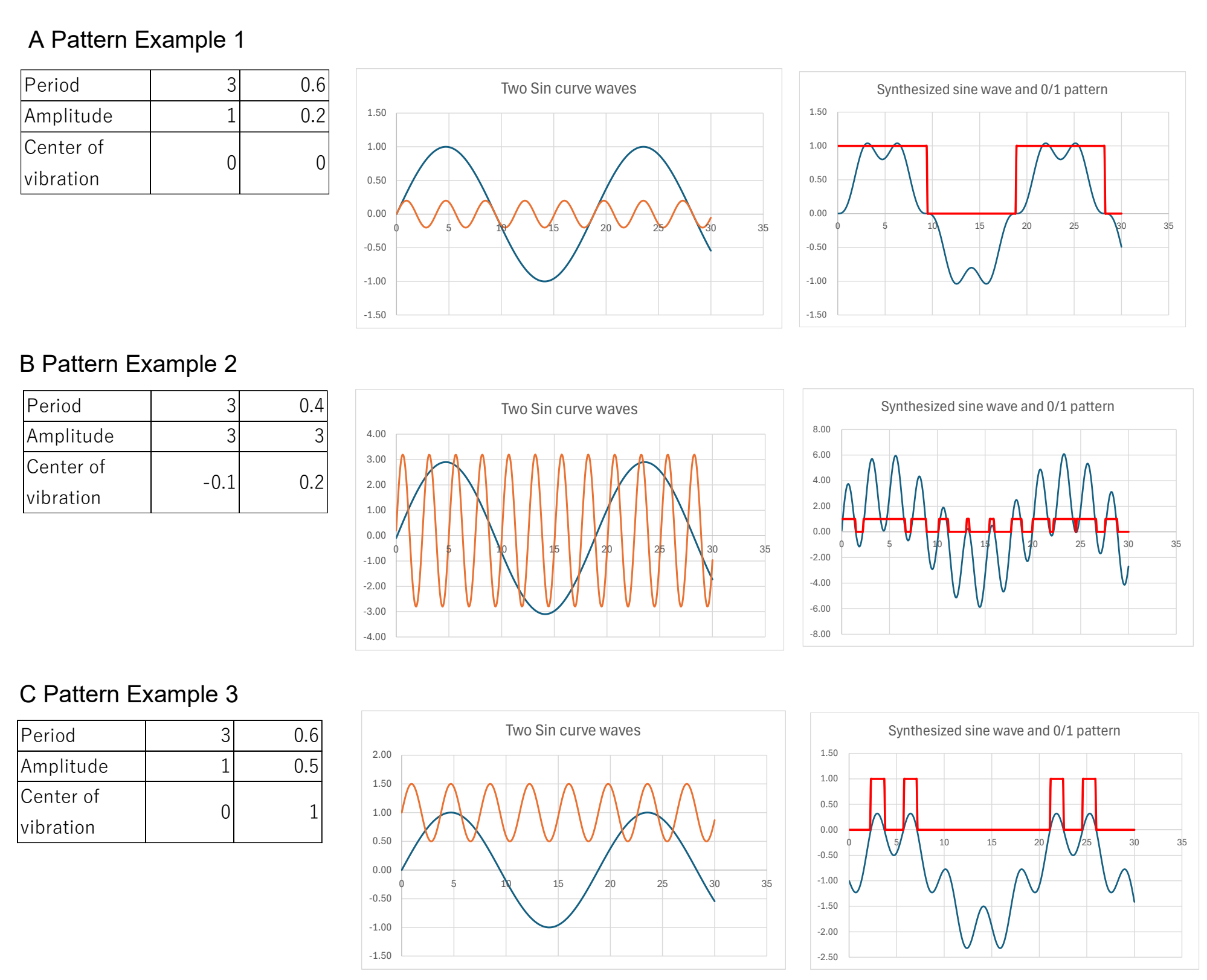

A Pattern Example 1

| Period | 3 | 0.6 |
|---|---|---|
| Amplitude | 1 | 0.2 |
| Center of vibration | 0 | 0 |

B Pattern Example 2

| Period | 3 | 0.4 |
|---|---|---|
| Amplitude | 3 | 3 |
| Center of vibration | -0.1 | 0.2 |

C Pattern Example 3

| Period | 3 | 0.6 |
|---|---|---|
| Amplitude | 1 | 0.5 |
| Center of vibration | 0 | 1 |

**Figure D1.** Examples of binary (0/1) patterns generated from the sign of the superposed values of two sine waves.

## Appendix E
## Examples of Binary (0/1) Patterns Generated from the Superposition of Two Lotka-Volterra Oscillations

A simulator was developed to demonstrate the interference of dual Lotka-Volterra oscillators and the subsequent generation of bit sequences. Example composite waveforms are presented below. Figure E1 shows the simulator interface and the waveforms of the two Lotka-Volterra oscillations. Figure E2 displays the 0/1 patterns (represented by the green lines in the graphs) generated from the composite waveform of these two oscillations using a threshold. It is demonstrated that a wide variety of 0/1 patterns can be generated by varying the Threshold (Bias).

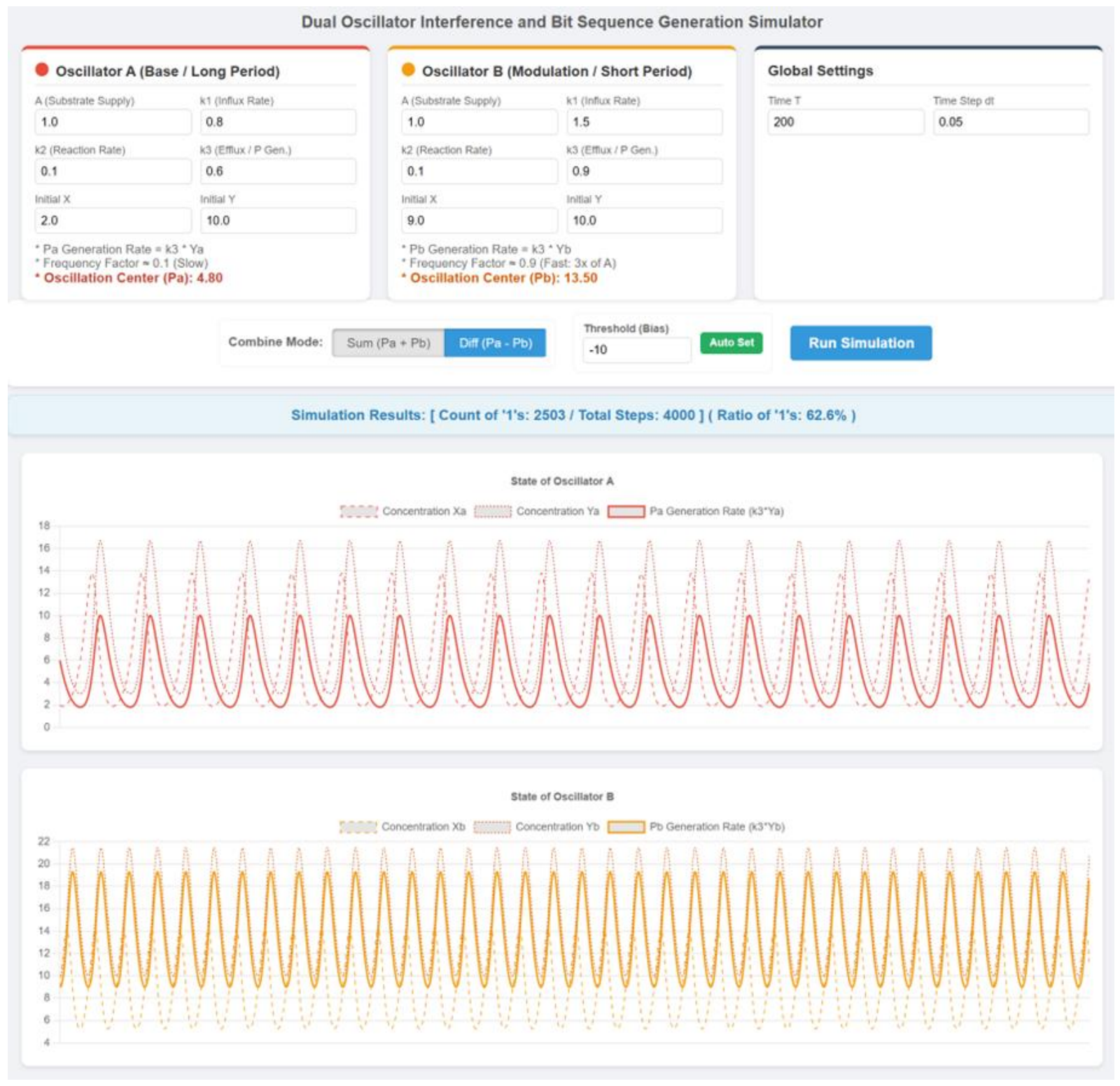

**Figure E1.** Simulator interface and waveforms of two Lotka-Volterra oscillations.

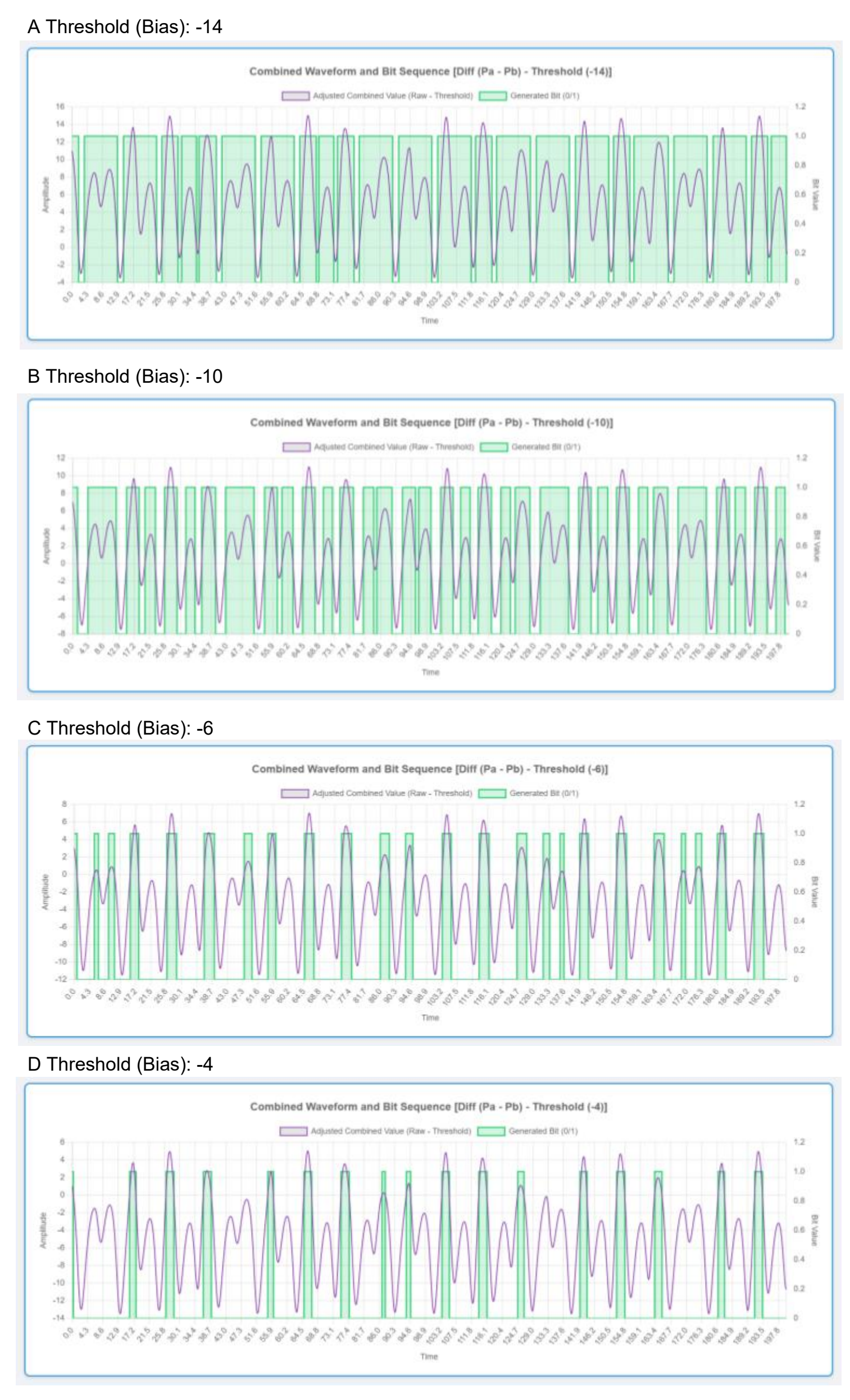

**Figure E2.** Variation of 0/1 patterns due to differences in the Threshold (Bias).

## Appendix F
## Sensitivity Analysis of Parameters

**(1) Polymerization Timing: dt (Lotka-Volterra Integration Step). Reference value: 0.08**

The success rates for *dt* ranging from 0.02 to 0.14 are summarized in Table F1. The setting of *dt* alters the interval of oscillation increments, thereby changing the timing of polymerization based on the 0/1 patterns. In this model, *dt* functions not merely as a measure of numerical precision but as a parameter directly linked to the "sampling interval of bits from the waves." There appears to be a "sweet spot" where the oscillation period and the 9-bit target sequence length resonance (synchronize), significantly increasing the probability of emergence. The results in the table suggest that the success rate of the catalytic loop exhibits a peak depending on the value of *dt*.

**Table F1.** Success Rate of Catalytic Loop Emergence with Different Time Steps (*dt*) (Random Seed 20 [5–34])

| dt | 0.02 | 0.04 | 0.06 | 0.08 (Ref) | 0.10 | 0.12 | 0.14 |
|---|---|---|---|---|---|---|---|
| Cognitive Model | 26.7 % | 43.3 % | 56.7 % | 83.3 % | 86.7 % | 73.3 % | 83.3 % |

**(2) Balance between Polymerization and Decomposition**

The results for variations in the natural polymerization rate, initial natural decomposition rate, incremental rate of natural decomposition (per step), and mutation rate (during replication) are shown in Tables F2 through F5.

Table F2 presents the success rates for different natural polymerization rates. While a higher polymerization rate increases the likelihood of generating catalytic sequences, success rates tend to decline when the rate becomes excessively high. This suggests that the timing and ratio of polymerization are critical factors influencing the success rate, warranting more detailed future investigations.

Table F3 shows the success rates for different initial natural decomposition rates. It is observed that the success rate slightly decreases even when the decomposition rate is low. This may imply that a very low decomposition rate reduces the opportunities for new sequences to emerge.

Table F4 shows the success rates for different incremental rates of natural decomposition (per step). Data indicating that excessively high incremental rates lead to the collapse of the established system represent the "error catastrophe" (limit of error). This can be interpreted as the limit of environmental harshness under which life can be sustained. Conversely, the success rate may also decrease if the incremental rate is too low, suggesting that a certain degree of decomposition may be beneficial.

Table F5 summarizes the success rates for different mutation rates. It is observed that higher mutation rates lead to higher success rates within the tested range.

**Table F2.** Success Rate for Different Natural Polymerization Rates (Random Seed 20 [5–34])

| Natural Polymerization Rate | 0.02 | 0.04(Ref) | 0.06 | 0.08 |
|---|---|---|---|---|
| Cognitive Model | 33.3 % | 83.3 % | 76.7 % | 63.3 % |

**Table F3.** Success Rate for Different Initial Natural Decomposition Rates (Random Seed 20 [5–34])

| Initial Natural Decomposition Rate | 0.06 | 0.10 | 0.14(Ref) | 0.18 | 0.20 |
|---|---|---|---|---|---|
| Cognitive Model | 63.3 % | 70.0 % | 83.3 % | 6.7 % | 0.0 % |

**Table F4.** Success Rate for Different Incremental Rates of Natural Decomposition (Random Seed 20 [5–34])

| Incremental Rate of Natural Decomposition | 0.0000002 | 0.000002 (Ref) | 0.00002 | 0.0002 | 0.002 | 0.02 |
|---|---|---|---|---|---|---|
| Cognitive Model | 70.0 % | 83.3 % | 70.0 % | 43.3 % | 6.7 % | 0.0 % |

**Table F5.** Success Rate for Different Mutation Rates (Random Seed 20 [5–34])

| Mutation Rate | 0.0001 | 0.001 (Ref) | 0.01 | 0.1 |
|---|---|---|---|---|
| Cognitive Model | 40.0 % | 83.3 % | 80.0 % | 90.0 % |

**(3) Parameters Related to Oscillator Settings**

Since changing the initial values of the Lotka-Volterra oscillations significantly alters the waveforms and often fails to produce suitable 0/1 patterns, the initial parameters of the oscillations were kept constant.

The results for variations in the composite wave threshold theta (Bias) and the substrate adjustment unit (the amount of fluctuation based on the difference between *Mplus* and *Mzero*) are presented. The "substrate adjustment unit" represents the step size for moving the oscillation center and is a critical parameter of the cognitive model, equivalent to the "learning rate" in artificial intelligence. When variations were tested across different orders of magnitude around 0.001, it was observed that excessively large values caused the center to shift too sensitively and erratically, leading to pattern instability (failure of learning). Conversely, excessively small values resulted in waveform changes that were too slow for the learning process to keep pace with natural decomposition (selection pressure), thereby reducing the success rate. These findings suggest that an appropriate learning speed is essential for emergence.

**Table F6.** Success Rate for Different Composite Wave Thresholds (Random Seed 20 [5–34])

| Threshold theta | -9 | -6(Ref) | -3 | 0 | 3 |
|---|---|---|---|---|---|
| Cognitive Model | 60.0 % | 83.3 % | 13.3 % | 0.0 % | 0.0 % |

**Table F7.** Success Rate for Different Substrate Adjustment Units (Random Seed 20 [5–34])

| Substrate Adjustment Unit | 0.0001 | 0.001 (Ref) | 0.01 | 0.1 |
|---|---|---|---|---|
| Cognitive Model | 46.7 % | 83.3 % | 0.0 % | 0.0 % |

**(4) Settings Related to Catalysts and tRNAs**

Tables F8 and F9 show the results when the weights for the Hamming distances of catalysts and tRNAs were varied. Naturally, larger weights facilitate the generation of catalysts. Conversely, smaller weights lead to lower success rates. These results serve as an indicator of the degree of "natural tolerance (slack)" required to overcome the "evolutionary valley."

**Table F8.** Success Rate of Catalytic Loop Emergence for Different Catalyst Hamming Distance Weights (Random Seed 20 [5–34])

| Catalyst Hamming Distance Weights | | | | |
|---|---|---|---|---|
| Distance 1 | 0.05 | 0.1 | 0.2 | 0.3 |
| Distance 2 | 0.025 | 0.05(Ref) | 0.1 | 0.15 |
| Cognitive Model | 6.7 % | 83.3 % | 73.3 % | 76.7 % |

**Table F9.** Success Rate of Catalytic Loop Emergence for Different tRNA Hamming Distance Weights (Random Seed 20 [5–34])

| tRNA Hamming Distance Weights | 0.005 | 0.01 (Ref) | 0.02 | 0.05 |
|---|---|---|---|---|
| Cognitive Model | 93.3 % | 83.3 % | 73.3 % | 70.0 % |

**(5) Settings Related to Protocell Evolution**

The number of protocells corresponds to the "population size" in genetic algorithms. If the number of cells is too small, diversity is lost, leading to stagnation where only inferior cells remain. Conversely, too many cells increase the computational load. This provides an opportunity to consider the "minimum population (society) scale required to secure diversity and pick up advantageous mutations." Success rates and evolutionary speeds were compared for 10, 20, 50 (reference), 80, and 120 protocells.

**Table F10.** Success Rate of Catalytic Loop Emergence for Different Protocell Numbers (Random Seed 20 [5–34])

| Number of Protocells | 10 | 20 | 50 (Ref) | 80 | 120 |
|---|---|---|---|---|---|
| Cognitive Model | 66.7 % | 93.3 % | 83.3 % | 83.3 % | 70.0 % |

The results for varying the weights of the fitness calculation formula are shown in Table F11. These weights represent "environmental demands." It is observed that setting the weight for tRNA (translation) extremely higher than for c60 (replication) can lead to a "local optimum (evolutionary dead-end)" where translation occurs but replication does not (preventing cell proliferation), resulting in a decrease in the success rate.

**Table F11.** Success Rate for Different Fitness Weighting Schemes (Random Seed 20 [5–34])

| Category | Reference | RNA Emphasis | Catalyst Emphasis | Codon Emphasis |
|---|---|---|---|---|
| c60/c61 (Replication/Separation) | 20 | 20 | 20 | 20 |
| tRNA-1/tRNA-2 (Translation Mach.) | 15 | 30 | 15 | 15 |
| c0/c1/c3 (Polym. Catalysts) | 10 | 10 | 30 | 10 |
| Codons (Memory) | 5 | 5 | 5 | 30 |
| Translation Bonus | 2 | 2 | 2 | 2 |
| Pool Max Length Bonus | 1 | 1 | 1 | 1 |
| Success Rate (Cognitive) | 83.3% | 66.7% | 73.3% | 63.3% |